\documentclass[reprint, amsmath, amssymb,aps,prl,floatfix, longbibliography]{revtex4-2}

\pdfoutput=1

\usepackage{amsmath,graphicx,natbib,bm,array,physics,hyperref}
\hypersetup{breaklinks=true, colorlinks=true, allcolors = blue}

\DeclareMathOperator{\sinc}{sinc}

\begin{document}

\newcommand{\pd}{\partial}
\newcommand{\beq}{\begin{equation}}
\newcommand{\eeq}{\end{equation}}
\newcommand{\bseq}{\begin{subequations}}
\newcommand{\eseq}{\end{subequations}}
\newcommand{\bpmat}{\begin{pmatrix}}
\newcommand{\epmat}{\end{pmatrix}}
\newcommand{\bpl}{\boldsymbol{(}}
\newcommand{\bpr}{\boldsymbol{)}}

\newcommand{\param}{\lambda}
\newcommand{\lloc}{L_{\text{loc}}}

\title{Scattering expansion for localization in one dimension}

\author{Adrian B. Culver}
\email{adrianculver@physics.ucla.edu}
\author{Pratik Sathe}
\author{Rahul Roy}
\email{rroy@physics.ucla.edu}
\affiliation{Mani L. Bhaumik Institute for Theoretical Physics, Department of Physics and Astronomy, University of California, Los Angeles, Los Angeles, CA 90095}

\date{\today}

\begin{abstract}
We present a perturbative approach to disordered systems in one spatial dimension that accesses the full range of phase disorder and clarifies the connection between localization and phase information.  We consider a long chain of identically disordered scatterers and expand in the reflection strength of any individual scatterer. As an example application, we show analytically that in a discrete-time quantum walk, the localization length can depend non-monotonically on the strength of phase disorder (whereas expanding in weak disorder yields monotonic decrease).  More generally, we obtain to all orders in the expansion a particular non-separable form for the joint probability distribution of the transmission coefficient logarithm and reflection phase.  Furthermore, we show that for weak local reflection strength, a version of the scaling theory of localization holds: the joint distribution is determined by just three parameters.
\end{abstract}

\maketitle

\setcounter{footnote}{0} 
\footnotetext[0]{For clarification on how a wave interference effect such as Anderson localization (i.e., an effect not uniquely quantum) can be relevant to quantum computing, see~\cite{LloydQuantum1999,BhattacharyaImplementation2002,KnightQuantum2003,Perez-GarciaQuantum2015}.}

\emph{Introduction.}---The localization of waves by disorder (Anderson localization) is a topic of enduring interest due to the wide range of settings in which it occurs, including electron transport, classical optics, acoustics, and Bose-Einstein condensates~\cite{Abrahams502010}.  Progress in the general theory of localization, independent of model details or of physical realization, can have similarly broad implications.  Another setting for localization -- of recent interest as a potential quantum computing platform~\cite{ChildsUniversal2009,LovettUniversal2010,SinghUniversal2021} -- is the quantum walk~\cite{KempeQuantum2003,Venegas-AndracaQuantum2012,KadianQuantum2021}, which is a quantum version of the classical random walk.  Localization has been demonstrated in quantum walks both experimentally and theoretically~\cite{PeretsRealization2008,SchreiberDecoherence2011,CrespiAnderson2013,VatnikAnderson2017,TormaLocalization2002,KeatingLocalization2007,YinQuantum2008,JoyeDynamical2010,AhlbrechtDisordered2011,ObuseTopological2011,VakulchykAnderson2017,DerevyankoAnderson2018} and could impact quantum computing proposals even in the idealized limit of no decoherence~\cite{KeatingLocalization2007,YinQuantum2008,ChandrashekarLocalized2015,Note0}.
    
A distinctive feature of localization in quantum walks is the prominent role of phase disorder.  Modern experimental platforms allow a high degree of control over a spatially-varying phase which can be disordered~\cite{PeretsRealization2008,SchreiberDecoherence2011,CrespiAnderson2013,VatnikAnderson2017}.  Localization in what is perhaps the simplest quantum walk -- a discrete-time quantum walk (DTQW) in one spatial dimension -- has been experimentally realized both for strong phase disorder~\cite{SchreiberDecoherence2011} and for a controllable range of phase disorder from weak to strong~\cite{VatnikAnderson2017}.   However, existing analytical approaches seem to apply only in the limiting cases when phase disorder is either weak or strong~\cite{VakulchykAnderson2017,DerevyankoAnderson2018}.  Furthermore, there are several phases that can appear in the quantum ``coin'' (see below) of a DTQW~\cite{ChandrashekarOptimizing2008,*ChandrashekarErratum2010,VakulchykAnderson2017}, and a localization calculation that allows them all to be disordered simultaneously seems to be lacking in the literature. 

In this Letter, we present a perturbative approach to localization in one spatial dimension.  Our approach accesses the full range of phase disorder and clarifies the connection between localization and phase information more broadly.  We use a general scattering setup~\cite{AndersonNew1980} that is applicable to DTWQs~\cite{TarasinskiScattering2014} and beyond.  A central feature of our approach is the relation between the localization properties and the reflection phase of a disordered scattering region~\cite{AndersonNew1980,LambertPhase1982,LambertRandom1983}.  (This phase has been measured in a DTQW experiment~\cite{BarkhofenMeasuring2017}.)  We calculate the localization length and the probability distribution of the reflection phase, and we extend the scaling theory of localization~\cite{AbrahamsScaling1979} to include correlations between the reflection phase and the transmission coefficient.

We now summarize our approach and results in more detail.  We consider a disordered region consisting of many single-channel scatterers that are independently and identically disordered, and we expand in the magnitude of the reflection amplitude of any individual scatterer \footnote{As we comment on in the main body of this manuscript, some of the results of our expansion at leading order were obtained in an equivalent form by Schrader \textit{et al}.~\cite{SchraderPerturbative2004}.  Also, in a more restricted setting, a scattering expansion similar to ours was done in~\cite{AsatryanSuppression2007} (see~\cite{GredeskulAnderson2012} for a review); however, this work has an uncontrolled aspect (c.f. the comment below Eq. 2.19 in~\cite{GredeskulAnderson2012}) not present in our work.  For further details, see~\cite{CulverScattering2024}.}.  Our first main result is the expansion of the inverse localization length.  We construct this expansion recursively and show that all orders depend only on local averages (that is, disorder averages over any single site).  We obtain a similar expansion of the probability distribution of the reflection phase, and indeed use this expansion in calculating the localization length.

As an example application of our first result, we calculate the localization length analytically as a function of arbitrary phase disorder in a two-component DTQW in one dimension.  We verify that our result interpolates between known results for weak and strong disorder that were calculated without reference to scattering~\cite{VakulchykAnderson2017}, and we find that the localization length can depend non-monotonically on the strength of phase disorder (similar to behavior seen numerically in~\cite{VatnikAnderson2017,VakulchykAnderson2017}) \footnote{As we discuss in~\cite{CulverScattering2024}, non-monotonicity in disorder strength has been found previously in particular cases of a particle scattering on random potential.  Also, it has been shown that transmission through a disordered region can be non-monotonic in disorder strength when the incident wave is in an energy gap of the non-disordered system (\cite{HeinrichsEnhanced2008} and references therein); we do not consider this case.}.  Our expansion applies when the quantum ``coin'' is highly biased (see below), which is a regime of interest for optimizing quantum search~\cite{ChandrashekarOptimizing2008,*ChandrashekarErratum2010}.  Even if the coin is only moderately biased, we find favorable agreement with numerics using the first two non-vanishing orders of our expansion.

Our second main result concerns the joint probability distribution $P_N(-\ln T,\phi)$, where $T$ is the transmission coefficient, $\phi$ the reflection phase, and $N$ the length of the disordered region.  We use an ansatz to find that for large $N$ and to all orders in the scattering expansion, $P_N(-\ln T,\phi)$ tends to a Gaussian function (of $-\ln T$) with mean, variance, and overall scale all depending on $\phi$ and all calculable order by order in terms of local averages.  We further show that at the leading order in the local reflection strength, a version of the scaling theory of localization applies: the joint distribution is determined by three parameters, which we may take to be the mean of $-\ln T$ and the mean and variance of $\phi$ \footnote{Recall that the scaling theory asserts (and theorems show~\cite{BougerolProducts1985}) that for large $N$, the marginal distribution $P_N(-\ln T)$ is Gaussian and is thus determined by two parameters: the mean and variance, both of which grow linearly with $N$~\cite{AbrahamsScaling1979,ShapiroScaling1987,CohenUniversal1988}.  The slope of the mean in $N$ is $2/\lloc$, defining the localization length $\lloc$ (sometimes defined without the $2$).  In certain regimes, the two parameters are related by an equation, i.e., ``single-parameter scaling'' holds~\cite{AbrahamsScaling1979,AndersonNew1980,DeychSingle2000, DeychSingleparameter2001,SchraderPerturbative2004}.}.  The latter two reach constant values for large system size.

In a companion paper~\cite{CulverScattering2024}, we present further discussion of the scaling theory and details of our calculations below.  We also present more applications, including a higher-order calculation in the Anderson model (yielding the leading dependence on the skewness of the on-site energy distribution) and results for a quantum particle scattering on a broad class of periodic-on-average random potentials (including as a special case the ``transparent mirror'' effect~\cite{BerryTransparent1997} from classical optics). 

\begin{figure}[htp]
    \includegraphics[scale=0.3]{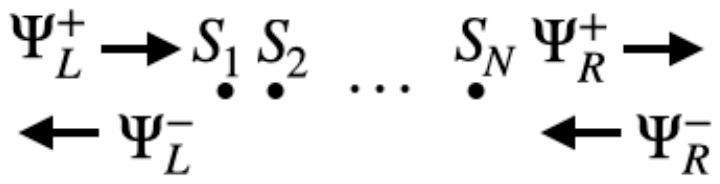}
    \caption{Schematic of our setup.}\label{fig:setup_diagram}
\end{figure}

\emph{Setup.}---We consider a general model of scattering through a disordered region (Fig.~\ref{fig:setup_diagram}).
The region consists of $N$ sites labeled as $n=1,\dots,N$, where each site $n$ is associated with a unitary $S$ matrix $\mathcal{S}_n$ parametrized as
\beq
    \mathcal{S}_n = 
    \bpmat
        t_n & r_n'\\
        r_n & t_n'
    \epmat,\label{eq:Sn general parametrization amplitudes}
\eeq
where $t_n$ and $t_n'$ ($r_n$ and $r_n'$) are the local transmission (reflection) amplitudes.  We consider only the single-channel case, i.e., these amplitudes are complex numbers and not matrices.    We take the disorder distribution of the S matrices to be independently and identically distributed (i.i.d.) across the $N$ sites; correlation between the entries of each individual $\mathcal{S}_n$ is allowed as long as every site has the same distribution.

The $S$ matrix for the region is obtained in the usual way by multiplying transfer matrices and is parametrized as in~\eqref{eq:Sn general parametrization amplitudes}, with (e.g.) $t_{1\dots N}\equiv \sqrt{T_{1\dots N}}e^{i\phi_{t_{1\dots N}}}$ and $r_{1\dots N}'\equiv \sqrt{R_{1\dots N}}e^{i\phi_{r_{1\dots N}'}}$.  We define $s_{1\dots N}= -\ln T_{1\dots N}$ for convenience, and we write the joint probability distribution of $s$ and $\phi_{r'}$ for the region as $P_N(s,\phi_{r'}) \equiv \langle \delta(s - s_{1\dots N})\delta(\phi_{r'} - \phi_{r_{1\dots N}'})\rangle_{1\dots N}$, where angle brackets indicate disorder averaging over the site or sites listed in the subscript.  (Except in $P_N$, we use subscripts to indicate dependence on the disorder parameters of the corresponding site or sites.)  Our task is to determine properties of $P_N(s,\phi_{r'})$, including the localization length (a property of the marginal distribution of $s$~\cite{Note3}), given the disorder distribution of the parameters of the local $S$ matrix~\eqref{eq:Sn general parametrization amplitudes}.

A basic assumption of our calculation is that localization occurs: that is, for large $N$ the region reflection coefficient $R_{1\dots N}\approx 1$ in all disorder realizations \footnote{Whenever we require a property to hold for all disorder realizations, we expect that it would suffice for that property to hold for almost all realizations (i.e., all except a set of measure zero).}.  The well-known exact recursion relations that determine $s_{1\dots N+1}$ and $\phi_{r_{1\dots N+1}'}$ from $s_{1\dots N}$, $\phi_{r_{1\dots N}'}$, $r_{N+1}$, and $r_{N+1}'$ then simplify for large $N$ to
\bseq
\begin{align}
    s_{1\dots N+1} &= s_{1\dots N} + g_{N+1}(\phi_{r_{1\dots N}'}),\label{eq:s recursion Rto1}\\
    \phi_{r_{1\dots N+1}'}&= \phi_{r_{1\dots N}'} + h_{N+1}(\phi_{r_{1\dots N}'}) \qquad(\text{mod } 2\pi),\label{eq:phirp recursion Rto1}
\end{align}
\eseq
where $g_n(\phi) = -\ln T_n + \ln(1- r_n e^{i\phi} - r_n^* e^{-i \phi} +R_n)$, $h_n(\phi)= \pi -i\ln \left( \frac{1-r_n^* e^{-i \phi}}{1-r_n e^{i\phi} } \frac{r_n r_n'}{R_n}\right)$, $R_n=|r_n|^2=|r_n'|^2$, and  $T_n=1-R_n$.  Equations~\eqref{eq:s recursion Rto1} and~\eqref{eq:phirp recursion Rto1} are the starting point for our analytical work, though we use the exact recursion relations in our numerical checks.

Our scattering expansion consists of rescaling $r_n \to \param r_n$ and $r_n' \to \param r_n'$ \footnote{The natural small parameter $\param$ in a particular problem may be such that $|r_n|$ starts at linear order in $\param$ but also has higher-order corrections.  Our results are easily adapted to this more general case~\cite{CulverScattering2024}.} in Eq.~\eqref{eq:Sn general parametrization amplitudes} (with $t_n$ and $t_n'$ also rescaled to maintain unitarity), then expanding in the parameter $\param$ while simultaneously sending $N\to\infty$ in a $\param$-dependent way such that the system is always in the localized regime.  In particular, we suppose that for any fixed $\param >0$ there is some $N_0(\param)$ for which $R_{1\dots N}\approx 1$ for any $N\ge N_0(\param)$ in all disorder realizations \footnote{A similar point is mentioned in~\cite{CohenUniversal1988} (footnote 24).}, and we always work in the regime $\param>0$ and $N\ge N_0(\param)$.  Below, we suppress $\param$ and refer informally to an expansion in $|r_n|$.

\emph{Scattering expansion of the localization length.}---We start by expressing the localization length in terms of the limiting form $p_{\infty}(\phi_{r'})\equiv \lim_{N\to\infty}\int_0^\infty ds\ P_{N}(s,\phi_{r'})$ of the marginal distribution of the reflection phase \footnote{In~\cite{CulverScattering2024}, we show that this limit exists provided that localization occurs.}.  From Eq.~\eqref{eq:s recursion Rto1}, we see that for sufficiently large $N$, $\langle s_{1\dots N}\rangle_{1\dots N}$ increases by the same constant amount each time $N$ is increased by one: $\langle s_{1\dots N+1} \rangle_{1\dots N+1} - \langle s_{1\dots N} \rangle_{1\dots N} =  \int_{-\pi}^{\pi} d\phi\ p_{\infty}(\phi) \langle g_{N+1}(\phi) \rangle_{N+1}= 2/\lloc$ \footnote{An equivalent expression was obtained by Lambert and Thorpe in~\cite{LambertPhase1982,LambertRandom1983}.}, where $\lloc$ is (by definition) the localization length~\cite{Note3}.  There is in fact no dependence on the particular site $N+1$ because the (i.i.d.) disorder average can be done over any site $n$ \footnote{Whenever we write a disorder average over $n$, it is understood that $n$ is an arbitrary site.}.  Converting to Fourier space yields the following series expression for the inverse localization length in terms of the Fourier coefficients $p_{\infty,\ell}\equiv \int_{-\pi}^\pi \frac{d\phi_{r'}}{2\pi}e^{-i\ell\phi_{r'}}p_{\infty}(\phi_{r'})$ and the moments of $r_n$~\cite{CulverScattering2024}:
\beq
    \frac{2}{\lloc} = \langle -\ln T_n\rangle_n
    - 4\pi \text{Re}\left[ \sum_{\ell=1}^\infty \frac{1}{\ell}p_{\infty,-\ell} \langle r_n^\ell \rangle_n\right].\label{eq:lloc in terms of Fourier coefficients}
\eeq

Equation~\eqref{eq:lloc in terms of Fourier coefficients} recovers the uniform phase formula $2/\lloc = \langle -\ln T_n\rangle_n$~\cite{AndersonNew1980} in two non-exclusive special cases: (i) the local reflection phase is uniformly distributed independently of the local reflection coefficient (then $\langle r_n^\ell\rangle_n =0$ for $\ell>0$), or (ii) the reflection phase distribution of the region is uniform.  Case (i) is an example of strong phase disorder.  The difficulty of applying Eq.~\eqref{eq:lloc in terms of Fourier coefficients}, in the case that (i) does not hold, is that it has been shown in many examples that the reflection phase distribution can be non-uniform, and in general the distribution is only known numerically (although Schrader \textit{et al}.~\cite{SchraderPerturbative2004} calculated $p_{\infty,\pm1}$ in an equivalent form) \footnote{The reflection phase distribution, and a related phase distribution in real space approaches, has been studied in a variety of models (see~\cite{PendrySymmetry1994} and references therein, and also, e.g.,~\cite{DorokhovLocalization1982,LambertPhase1982,JayannavarPhase1990,JayannavarScaling1991,HeinrichsRelation2002,SchomerusBandcenter2003,TessieriAnomalies2018,PradhanPhase2021}).  Also, $p_\infty(\phi_{r'})$ is equivalent ~\cite{CulverScattering2024} to the ``invariant measure'' that has been studied and related to the Lyapunov exponent in the mathematics literature~\cite{FurstenbergNoncommuting1963,BougerolProducts1985,JitomirskayaDelocalization2003, SchraderPerturbative2004}.}.

The key advance that we make is to apply the scattering expansion to $p_{\infty}(\phi_{r'})$, showing that its Fourier coefficients may be written as a recursively defined series involving only \emph{local} averages.  Our calculation relies on the disorder distribution being ``reasonable'' and the particular model parameters chosen being ``generic;'' our precise assumptions are that localization occurs and that the inequality $\left\langle e^{i\ell(\phi_{r_n}+\phi_{r_n}'+\pi)}\right\rangle_n \ne 1$ holds for all integers $\ell\ne 0$ \footnote{Violation of this inequality requires a particular alignment of the reflection phases across disorder realizations.  In~\cite{CulverScattering2024}, we consider the case that the inequality holds only for $0<|\ell|\le\ell_\text{max}$ and show that our expansion is valid up to an order that increases with $\ell_\text{max}$.  For instance, Eq.~\eqref{eq:lloc to 4th order} holds if $\ell_\text{max}\ge5$.}.

We focus here on the results of this calculation; see~\cite{CulverScattering2024} for details.  It is convenient to define $v_n= r_n r_n'/ R_n$, $\alpha_\ell = 1/[1 - \langle (-v_n)^\ell\rangle_n ]$, and several constants determined by local averages (we use a superscript to indicate the order of a given constant in the scattering expansion): $\gamma^{(1)} = \alpha_1 \langle r_n'\rangle_n$, $\gamma^{(2)} = \alpha_2 \langle r_n' (r_n'-2\gamma^{(1)}v_n)\rangle_n$, $\gamma_1^{(3)} =\alpha_1 \langle r_n  (\gamma^{(1)}r_n' - \gamma^{(2)}v_n)\rangle_n$, and $\gamma_3^{(3)} = \alpha_3 \langle r_n' ({r_n'}^2- 3\gamma^{(1)}r_n'v_n +3 \gamma^{(2)}v_n^2 )\rangle_n$.  Then we have $2\pi p_{\infty}(\phi_{r'}) = 1 + 2\text{Re}[ (\gamma^{(1)} + \gamma_1^{(3)})e^{-i\phi_{r'}} + \gamma^{(2)} e^{-2i\phi_{r'}} + \gamma_3^{(3)}e^{-3i\phi_{r'}}] + O(|r_n|^4)$ and our main result for the localization length:
\begin{widetext}
\begin{multline}
    \frac{2}{\lloc} = \langle R_n\rangle_n - 2 \text{Re}\left[\frac{\langle r_n\rangle_n\langle r_n'\rangle_n}{1 +\langle r_n r_n'/R_n\rangle_n}  \right]\\
    + \frac{1}{2}\langle R_n^2\rangle_n - \text{Re}\left[\alpha_2 ( \langle r_n^2\rangle_n -2\alpha_1\langle r_n\rangle_n \langle r_n v_n\rangle_n) ( \langle r_n'^2\rangle_n -2\alpha_1\langle r_n'\rangle_n \langle r_n' v_n\rangle_n)
    + 2\alpha_1^2\langle r_n\rangle_n \langle r_n'\rangle_n \langle r_n r_n'\rangle_n\right] + O(|r_n|^6).\label{eq:lloc to 4th order}
\end{multline}
\end{widetext}
The first two terms in Eq.~\eqref{eq:lloc to 4th order} are the leading-order contribution (second order in $|r_n|$) and were found in an equivalent form by Schrader \textit{et al}. in~\cite{SchraderPerturbative2004}.  The remaining terms are fourth order, and indeed all odd orders vanish by symmetry~\cite{CulverScattering2024}.  The terms whose real parts are taken are the contributions from the non-uniformity of the reflection phase distribution.  We emphasize that these non-uniform phase contributions are parametrically of the same order as the uniform phase contributions ($\langle -\ln T_n\rangle_n = \langle R_n\rangle_n + \frac{1}{2}\langle R_n^2\rangle_n +\dots$); in particular, deviations from phase uniformity generally affect the inverse localization length even at leading order~\cite{LambertRandom1983, JitomirskayaDelocalization2003, SchraderPerturbative2004}.

\emph{Application to quantum walks.}---We next apply the general result~\eqref{eq:lloc to 4th order} to a single-step, two-component DTQW in one dimension.  The setup is an infinite chain with site index $n$ and an internal ``spin'' degree of freedom ($\uparrow$ or $\downarrow$).  The unitary operator $\hat{U}$ that implements a single time step is $\hat{U}=\sum_n \left( \ket{n+1 ,\uparrow} \bra{n, \uparrow} + \ket{n-1,\downarrow}\bra{n,\downarrow} \right)\hat{U}_\text{coin}$, where the ``coin'' operator is $\hat{U}_\text{coin} = \sum_n \ket{n}\bra{n}\otimes U_{\text{coin},n}$ and $U_{\text{coin},n}$ is a general $2\times2$ unitary matrix (acting on the spin degree of freedom at site $n$) parametrized as~\cite{ VakulchykAnderson2017}
\beq
U_{\text{coin},n} =   e^{i\varphi_n} \bpmat
    e^{i\varphi_{1,n}} \cos \theta_n & e^{i\varphi_{2,n}} \sin\theta_n \\
    -e^{-i\varphi_{2,n}}\sin\theta_n & e^{-i\varphi_{1,n}}\cos\theta_n
\epmat.\label{eq:Ucoin,n}
\eeq
We take the parameters $\mathbf{D}_n\equiv (\varphi_n,\varphi_{1,n}, \varphi_{2,n},\theta_n)$ to be i.i.d. across the sites $n=1,\dots,N$ (note that the components of $\mathbf{D}_n$ may be correlated with each other), defining a disordered region.

The $S$ matrix of the region describes solutions of the stationary state equation $\hat{U}\ket{\Psi}=e^{-i \omega}\ket{\Psi}$, where $\omega$ is the quasienergy.  There are in fact many possible scattering problems, corresponding to different choices for site-independent values to be assigned to $\mathbf{D}_n$ in the non-disordered regions (the sites $n<1$ and $n>N$).  It may be shown that all choices result in a problem of the form we have been considering (i.e., there is some $S$ matrix $\mathcal{S}_n$ that depends only on $\mathbf{D}_n$ and $\omega$) and that the probability distribution of the transmission coefficient in the localized regime is the same in all cases~\cite{CulverScattering2024}.  We consider the simplest case of setting $\mathbf{D}_n=\mathbf{0}$ in the non-disordered regions~\cite{TarasinskiScattering2014}, which results in $\mathcal{S}_n = e^{i\omega}U_{\text{coin},n}$.  Comparing to Eq.~\eqref{eq:Ucoin,n}, we see that the local reflection amplitudes are $r_n = -e^{i(\omega+\varphi_n - \varphi_{2,n})}\sin \theta_n$ and $r_n' = e^{i(\omega+\varphi_n + \varphi_{2,n})}\sin \theta_n$.  Then Eq.~\eqref{eq:lloc to 4th order} yields the inverse localization length for small $\sin\theta_n$, up to an error of order $|r_n|^6=\sin^6\theta_n$, with arbitrary phase disorder.  In particular, the distribution of $\mathbf{D}_n$ is arbitrary as long as $\sin\theta_n$ is always small.

\begin{figure}[htp]
    \includegraphics[width=\columnwidth]{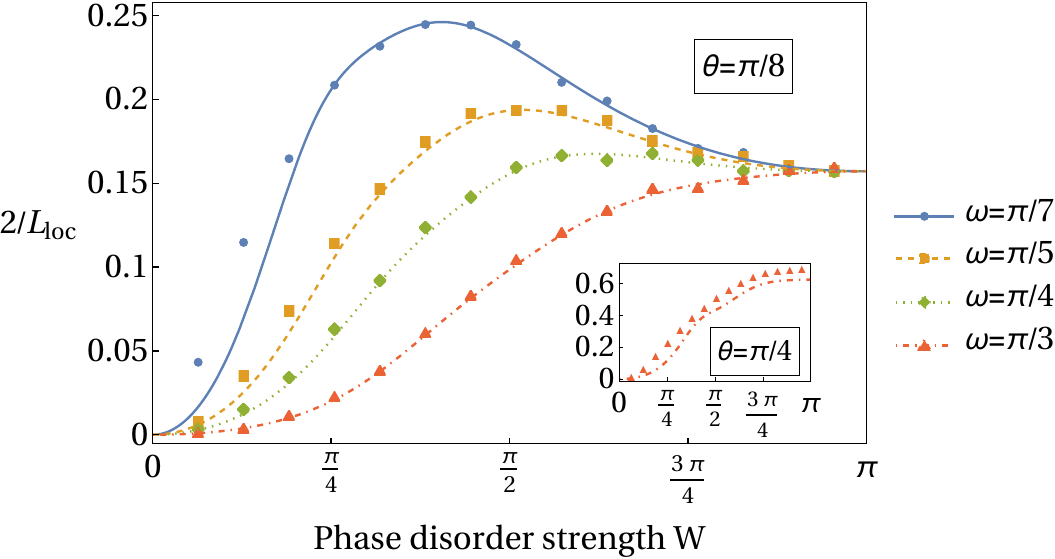}
    \caption{The inverse localization length ($2/\lloc$) vs. the strength of phase disorder ($W$) in the variable $\varphi_n$ in the DTQW.  We compare our theoretical result~\eqref{eq:lloc to 4th order} (lines) with numerics (points) for a moderately biased coin (main plot) and for an unbiased coin (inset).}\label{fig:larger_coin_parameters_plot}
\end{figure}

Specializing to the case of $\varphi_n$ uniformly distributed in $[-W,W]$, with $\varphi_{1,n}=\varphi_{2,n}=0$ and $\theta_n\equiv \theta$, we obtain the inverse localization length for small $\sin\theta$ and arbitrary phase disorder strength $W$.  We have verified that our result agrees with the calculation of Vakulchyk \textit{et al}.~\cite{VakulchykAnderson2017}, in which $\theta$ is arbitrary and $W$ is either small (yielding $2/\lloc\sim W^2$) or equal to $\pi$ (in which case the uniform phase formula holds).  Our result thus interpolates between the known limits of weak and strong phase disorder and analytically demonstrates non-monotonicity in disorder strength \footnote{$\frac{2}{\lloc} = \left( 1 + \frac{2\sinc^2W[\cos(2w)-\sinc(2W)]}{1-2\cos(2w)\sinc(2W)+\sinc^2(2W)}\right)\sin^2\theta$ (where $\sinc x= \frac{\sin x}{x}$) is the explicit result at leading order and exhibits non-monotonicity in $W$ for $\omega$ within a certain range.}.
We have verified our result with numerics in the regime of small $\sin\theta$~\cite{CulverScattering2024}; furthermore, in Fig.~\ref{fig:larger_coin_parameters_plot} we show that the agreement with numerics is favorable even if $\sin\theta$ is not particularly small.  

\emph{Joint probability distribution.}---Returning to the general case, we now summarize the results of applying the scattering expansion to the joint probability distribution $P_{N}(s,\phi_{r'})$~\cite{CulverScattering2024}.  We find that for large $N$ this distribution takes a Gaussian form defined as follows.  There is a constant $c$ and two functions $\hat{s}(\phi_{r'})$, $\eta(\phi_{r'})$ for which we have
\begin{multline}
    P_{N}(s,\phi_{r'}) = p_{\infty}(\phi_{r'})\\
    \times \frac{1}{\sqrt{2\pi \sigma(N,\phi_{r'})^2}} e^{-\frac{1}{2}[ s- \frac{2N}{\lloc} - \hat{s}(\phi_{r'})]^2/\sigma(N,\phi_{r'})^2},\label{eq:joint prob dist final}
\end{multline}
where the phase-dependent variance $\sigma(N,\phi_{r'})^2$ scales linearly with $N$ with a sub-leading, phase-dependent correction: $\sigma(N,\phi_{r'})^2 = 2 [c N + \eta(\phi_{r'})]$.  The constant $c$ is related to the variance $\sigma(N)^2$ of the marginal distribution of $s$ by $\sigma(N)^2 = 2cN + O(N^0)$.   We can calculate the quantities $c$, $\hat{s}(\phi_{r'})$, and $\eta(\phi_{r'})$ order by order in the scattering expansion in terms of local averages [except that the functions $\hat{s}(\phi_{r'})$ and $\eta(\phi_{r'})$ each have an undetermined, $\phi_{r'}$-independent additive constant], and in particular we have obtained $c= 2/\lloc + O(|r_n|^4)$ (which was shown in an equivalent form in~\cite{SchraderPerturbative2004} with a third-order error term), $\hat{s}(\phi) =  2\text{Re}\{ \gamma^{(1)}e^{-i\phi} + [\frac{3}{2}\gamma^{(2)} - (\gamma^{(1)})^2]e^{-2i\phi} \} + O(|r_n|^3)+\text{const.}$, and $\eta(\phi) =\text{Re}\{ [\gamma^{(2)} - (\gamma^{(1)})^2 ] e^{-2i\phi}] \} + O(|r_n|^3) + \text{const}$.

We now explain briefly how we arrive at Eq.~\eqref{eq:joint prob dist final}.  From Eqs.~\eqref{eq:s recursion Rto1} and~\eqref{eq:phirp recursion Rto1}, it is straightforward to show that the joint probability distribution satisfies a recursion relation of the form $P_{N+1}(s,\phi_{r'}) = \mathcal{F}[s,\phi_{r'};\{P_{N}\}]$, where $\mathcal{F}$ is a linear functional in its last argument.  We take Eq.~\eqref{eq:joint prob dist final} as an ansatz and require $\mathcal{F}[s,\phi_{r'};\{P_{N}\}]= P_{N+1}(s,\phi_{r'}) + O(1/N^2)$ for large $N$; this requirement fixes $c$, $\hat{s}(\phi_{r'})$, and $\eta(\phi_{r'})$ to all orders in the scattering expansion [except for the constant offsets of $\hat{s}(\phi_{r'})$ and $\eta(\phi_{r'})$].  Since the ansatz itself is $O(1/\sqrt{N})$, we can expect that~\eqref{eq:joint prob dist final} is the leading term in an expansion in $1/\sqrt{N}$ of the exact answer.

The correlation between $s$ and $\phi_{r'}$ in~\eqref{eq:joint prob dist final} is a finite-size effect, as we now explain.  We write the average of $s$ as $\langle s \rangle = 2N/\lloc + O(N^0)$, and we consider how accurate $\langle s\rangle$ is as an estimate of the conditional average of $s$ with fixed $\phi_{r'}$ in~\eqref{eq:joint prob dist final}.  The phase-dependent variation of the mean introduces a relative error of order $\hat{s}(\phi_{r'})/\langle s\rangle \sim1/N$, while the finite variance introduces a relative error of order $\sigma(N,\phi_{r'})/\langle s\rangle = c\lloc/\sqrt{N} + O(N^{-3/2})$, where the $N^{-3/2}$ term contains the contribution of the function $\eta(\phi_{r'})$.  Prior work has found the joint probability distribution to factorize into a transmission coefficient part times a phase part~\cite{RobertsJoint1992, PendrySymmetry1994}, in apparent contradiction to our Eq.~\eqref{eq:joint prob dist final}; this suggests that the prior work only accounted for the $1/\sqrt{N}$ term in the above discussion and neglected the $1/N$ and $N^{-3/2}$ terms that contain the  correlations between $s$ and $\phi_{r'}$.

We next show that the scaling theory applies to the joint distribution in the regime of weak local reflection strength.   Here we ignore $\eta(\phi_{r'}$) (whose effect is subleading for large $N$, as we have shown above) and expand the remaining terms of~\eqref{eq:joint prob dist final} to leading order in the scattering expansion.  A single parameter~\cite{Note3} suffices to determine $2/\lloc$ and $c$ since they are equal at leading order~\cite{SchraderPerturbative2004,CulverScattering2024}.  Furthermore, the phase distribution up to first order is determined entirely by two parameters: the real and imaginary parts of $\gamma^{(1)}$, or by a simple change of variables, the mean and variance of $\phi_{r'}$.  The key relation that implies that these three parameters suffice to determine the joint distribution is that the first-order part of the phase-dependent mean turns out to be essentially the same function as the first-order part of the phase distribution:
\beq
    \hat{s}(\phi_{r'}) = 2\pi p_{\infty}(\phi_{r'}) + O(|r_n|^2) + \text{const.},
\eeq
where the constant on the right-hand side is independent of $\phi_{r'}$.

\emph{Conclusion.}---In a general problem of single-channel scattering through an i.i.d. disordered region, we developed a systematic expansion in the local reflection strength, which we call the scattering expansion.  We calculated the inverse localization length to the first two non-vanishing orders in this expansion, using an explicit expansion of the (generally non-uniform) reflection phase distribution.  We applied our result to calculate the localization length in a two-component DTQW with a biased coin parameter and arbitrary phase disorder, and we thus showed analytically that the localization length can depend non-monotonically on the strength of phase disorder.

Returning to the general problem, we summarized the results of applying the scattering expansion to the joint probability distribution of the transmission coefficient logarithm and reflection phase: first, we found the general form of the joint distribution to all orders in the scattering expansion, and second, we showed that when the local reflection strength is weak, the joint distribution is determined by three parameters.

It would be interesting to explore implications that our scattering-based approach might have for the more usual DTQW setup, in which a walker starts in a spatially confined initial state and evolves in time.  Ballistic spread (i.e., variance increasing quadratically with time) is an important property of DTQWs and is known to be suppressed by localization.  However, if the localization length is sufficiently large, then this suppression would be unimportant, since the walker can be expected to travel ballistically until reaching a distance of order $\lloc$.  (It has indeed been found in a particular model that the maximum distance that can be reached by the walker has the same scaling with disorder strength as the localization length~\cite{KeatingLocalization2007}.)  Our scattering-based results for $\lloc$ might yield an upper bound (after appropriate maximization over quasienergy) on the $\lloc$ that appears in the time-dependent problem.  Also, our technique for calculating the reflection phase distribution might extend to the distribution of the Wigner delay time ($d\phi_{r'}/d\omega$), which would characterize the time that a walker spends in being reflected from a disordered region in an otherwise non-disordered environment.

Another direction to explore would be applications of our approach to other problems involving products of random matrices, even outside the setting of scattering theory.  For instance, in the study of randomly-driven conformal field theories, Ref.~\cite{WenPeriodically2022} encounters a problem that seems to fit our framework [a product of random SU$(1,1)$ matrices]; each matrix represents a time step, and the Lyapunov exponent (inverse localization length) is shown to be the rate of entanglement entropy growth (and to be a lower bound on the heating rate). 

Finally, this work could be a step towards an analytical treatment of the quasi-one-dimensional case (i.e., many scattering channels rather than one).  This would be significant because the quasi-one-dimensional case can be used to study delocalization transitions in dimensions higher than one; in particular, one studies (usually numerically) the scaling, as the number of transverse modes goes to infinity, of the largest localization length~\cite{RomerNumerical2022}.  If we can carry out our approach with multiple scattering channels, then the possibility could arise of taking this limit analytically.  This could provide a perturbative handle on critical exponents in higher-dimensional localization-delocalization transitions, such as the plateau transition in the integer quantum Hall effect~\cite{ChalkerPercolation1988}.  We note that Ref.~\cite{ChalkerScattering1993} finds in a particular model that departure from (multichannel) phase uniformity is necessary for obtaining a metal-insulator transition; this suggests that a quasi-one-dimensional version of our approach would be useful.

\emph{Acknowledgments.}---We thank Albert Brown and Fenner Harper for collaboration on related work, and we thank Victor Gurarie, Daniel Lidar, and Abhinav Prem for discussion.  A.B.C., P.S., and R.R. acknowledge financial support from the University of California Laboratory Fees Research Program funded by the UC Office of the President (UCOP), grant number LFR-20-653926.  A.B.C acknowledges financial support from the Joseph P. Rudnick Prize Postdoctoral Fellowship (UCLA).  P.S. acknowledges financial support from the Center for Quantum Science and Engineering Fellowship (UCLA) and the Bhaumik Graduate Fellowship (UCLA).  This work used computational and storage services associated with the Hoffman2 Shared Cluster provided by UCLA Institute for Digital Research and Education’s Research Technology Group.  

\bibliography{ms.bib}

\end{document}